\documentclass[11pt]{JHEP3}

\usepackage{epsfig}
\usepackage{graphicx}

\preprint{UOSTP-03-105\\SNUTP 03-022\\{\tt hep-th/0310123}}

\title{\large \bf Troubles with Spacetime Noncommutative Theories:
Tachyons or S-branes? }

\author{Dongsu Bak \\
\\
Physics Department, University of Seoul,
Seoul 130-743, Korea \\
dsbak@mach.uos.ac.kr}

\author{Seok Kim \\
\\
School of Physics, Seoul National University, Seoul 151-747, Korea \\
calaf2@snu.ac.kr}

\abstract{We find Lorentzian solutions of spacetime noncommutative
gauge theories that are localized exponentially in space and time.
Together with time translational invariance of the theories,
 we argue that perurbative S matrix formulation of
such theories is problematic in the sense that the S matrix based
on free in and out states misses the spacetime localized degrees.
We show that, in 3+1 dimensions, the problem disappears for the
cases where the noncommutativity becomes purely spatial by an
appropriate Lorentz transformation or the  noncommutativity is
lightlike with its electric and magnetic parts orthogonal to each
other.}

\begin{document}

\def\nn{\nonumber}

\section{Introduction}

There have been some discussions about whether  spacetime
noncommutative theories are well defined as
 quantum theories\cite{gomismehen}-\cite{CH}. At first,
it was argued that spacetime noncommutative theories have a
unitarity problem by studying
the 
S-matrices via the perturbative
analysis\cite{gomismehen,aha,al}. 
This is consistent with the fact that there is no decoupling limit
from string theories leading to the spacetime noncommutative field
theories for specific ranges of the noncommutativity
parameter\cite{gomismehen,aha}.
 Later
it was claimed that one may recover the perturbative unitarity
 by the careful
treatment of time orderings\cite{bahns,rim}. There have been
further studies\cite{torr,ohl} in this direction but the status of
the matter seems not  clear at this moment.

The spacetime noncommutative field theories involve infinitely
many time derivatives. There is no well established procedure by
which one can define quantization
of 
such theories. In the perturbative analysis, Feynman rules are
given as a definition of the quantum theory
but its justification
 is never clear.

In this note, we shall consider spacetime noncommutative field
theories only classically. By solving equations of motion with
Lorentzian signature, we shall construct solutions that are
exponentially localized in time as well as spatial directions. Due
to the translational invariance of the theory, one may find such
solutions centered at any points in spacetime once there exist
any. The existence of such solution implies that arbitrarily small
fluctuations in the far past may become nontrivial ones of order
one in the far future
when such objects are translated  to the far future. 
To put an emphasis on this, we shall show that such spacetime 
localized degrees may arise not just from the free vacuum but also
from  scattering states by explicitly constructing solutions of
spacetime localized degrees superposed with scattering states.
Perturbative S matrix formulation of the spacetime noncommutative 
theory will  be
problematic because the S-matrix based on free part degrees of
freedom 
would miss the spacetime localized
degrees.
More precisely,  scattering states alone and scattering states with 
spacetime localized degrees are 
physically distinct but 
cannot be distinguished, in the far past, as an initial data. 
Nonperturbative treatment of the system may cure the problem but
we do not know how to do this.


We shall illustrate existence of spacetime localized solutions by
considering various noncommutative gauge theories in 1+1 and 3+1
dimensions for certain choice of noncommutative parameters
$\theta^{\mu\nu}$ with nonvanishing electric components. When the
noncommutativity is
 purely spatial, it is quite clear that there is no problem
of unitarity at least classically and   the standard procedure of
quantization works
 without any troubles.
As in \cite{gomismehen,aha}, Lorentz invariant combinations of
$\theta^{\mu\nu}$ are relevant here. Specifically for 3+1
dimensions, the cases of $\vec{\theta}_e \cdot \vec{\theta}_m=0$
and
$\vec{\theta}_e^2 \le \vec{\theta}_m^2$ with $\theta_e^i=
\theta^{0i}$ and $\theta_m^i={1\over 2} \epsilon^{ijk}\theta^{jk}$
will be free of problems. Otherwise the gauge theories
are plagued with the 
Lorentzian solutions that are localized both in space and time.
For the $\theta^{\mu\nu}$
 without such problematic solutions,
string theories in the NS-NS B-field background allow the
decoupling field-theory limit\cite{gomismehen,aha}. Therefore we
get an agreement.

All of these solutions are closely related to the solitons,
vortices, and instantons in the noncommutative gauge theories with
spatial
noncommutativity\cite{polychronakos,bak,aganagic,harvey,gross,klee,yee}.
With purely spatial noncommutativity, the string theory
interpretation of the solution is known. For example, the vortex
solution in 2+1 dimensional U(1) gauge theory corresponds to
unstable D0 in D2 with NS-NS B fields\cite{aganagic}. In the
present
 case, we do not find such an
interpretation. If they were  Euclidean solutions, one might
interpret them as instantons, but here we are dealing with
Lorentzian solutions.

In Section 2, we shall construct the spacetime localized solutions
in 1+1 and 3+1 dimensional gauge theories interacting with scalars
on various noncommutative spacetimes.
 We won't try to find all of such solutions
because finding only one will be enough to say that the theory is
problematic.
For U(1) or U(N) gauge theories\cite{park}, one may construct such
solutions once the conditions on $\theta^{\mu\nu}$ are met.
Section 3 will be devoted to discussions. We  give an
interpretation of the solutions and  comment on the spacetime
noncommutative scalar theories.

\section{Spacetime Localized Lorentzian Solutions}

We will present various theories and solutions in terms of
ordinary functions equipped with the star product
\begin{equation}\label{star}
  f(x)\ast g(x)\equiv
  \left.
  e^{\frac{i}{2}
\theta^{\mu\!\nu}\partial_{\mu}\partial_{\nu}^{\prime}}f(x)g(x')\
  \right|_{x=x'}\ .
\end{equation}
There is a one-to-one correspondence between
 ordinary function and  operator, where
operators 
act on
the Hilbert space
built upon  the commutation relations,
\begin{equation}
  [\hat x^{\mu},\hat x^{\nu}]=i\theta^{\mu\nu}\ .
\end{equation}
With the map realizing the correspondence, 
\begin{equation}
  \hat f(\hat x) =\int \frac{d^{D}\!k\
  d^{D}\!x}{(2\pi)^{D}}f(x)e^{ik\cdot (\hat x - x)}\ ,
\end{equation}
 the operator product can be translated to the star product in
the function space.

Let us consider the 1+1 dimensional
spacetime noncommutative U(1) gauge theory, 
\begin{equation}\label{2d-gauge}
  S=-\frac{1}{4g^2}\int d^2x\, F_{\mu\nu}\!\ast\!F^{\mu\nu}
\end{equation}
where
\begin{equation}
  F_{\mu\nu} =
  \partial_{\mu}A_{\nu}-\partial_{\nu}A_{\mu}-
i(A_{\mu}\!\ast\!A_{\nu}-A_{\nu}\!\ast\!A_{\mu})\ .
\end{equation}
We introduce here the  spacetime noncommutativity, $[\hat x^1,
\hat x^0]=-i\theta$, where we take $\theta >0$ without loss of
generality. The equation of motion reads
\begin{equation}\label{u(1)-eq-general}
  D_{\mu}F^{\mu\nu}=0\,,
\end{equation}
or explicitly
\begin{equation}\label{u(1)-eq}
  D_{0}F_{01}=D_{1}F_{01}=0\ .
\end{equation}
A class of solutions of (\ref{u(1)-eq}) may be obtained by
recalling some 
static
solutions of 2+1 dimensional 
U(1) gauge theory 
on a spatially noncommutative plane $[\hat x^1,\hat
x^2]=-i\theta$. The equation of motion satisfied by static
configurations is $D_{1}F_{12}=D_{2}F_{12}=0$, which becomes
exactly the same as (\ref{u(1)-eq}) by replacing $x^{2}\rightarrow
x^{0}$ and $A_{2}\rightarrow A_{0}$. (Note that the equation
(\ref{u(1)-eq}) is
 insensitive to the
signature of the metric in two dimensions.) In
Ref.\cite{polychronakos, bak},  solutions carrying magnetic flux
are constructed. In  the operator form,  the solution reads
\begin{equation}\label{potential}
  \hat A_{i}=-\frac{1}{\theta}\epsilon_{ij}
(\hat x^{j}-\hat S\hat x^{j}\hat S^{\dag})\,, \ \ \ (i,j=1,2)
\end{equation}
where the shift operator $\hat{S}$ is defined by
\begin{equation}
\hat S^\dag\hat S=1,\ \ \ \   \ \hat S \hat S^\dag=1-\hat P,
\end{equation}
with  
 any projection operator $\hat P$  satisfying
$\hat P^2=\hat P$. The field strength is
\begin{equation}\label{b-field}
  \hat F_{12}=\frac{1}{\theta}\hat{P}.
\end{equation}
For simplicity, we take the solution
$F_{12}=\frac{1}{\theta}|0\rangle\langle 0|$ with the choice
$S=\sum_{n=0}^\infty |n+1\rangle\langle n|$ and use it to get the
solution of the \textit{Lorentzian} theory with spacetime
noncommutativity. Mapping the solution  to function form,
one finds that the field strength is localized in spacetime  as
\begin{equation}\label{e-field}
  F_{10}= \frac{2}{\theta}\ \exp\left[-\frac{x^2+t^2}{\theta}\right]\ .
\end{equation}

One may find similar solutions for 1+1 dimensional Abelian-Higgs
theory on the noncommutative spacetime.
Adding a
fundamental scalar with quartic potential, 
the action for the Abelian-Higgs theory becomes
\begin{equation}\label{2d-gauge-fund}
S=  -\frac{1}{g^2}\int d^2x
\left(\frac{1}{4}F_{\mu\nu}\!\ast\!F^{\mu\nu}+
  D_{\mu}\phi\ast(D^{\mu}\phi)^{\dag}+
  \lambda(\phi\ast\phi^{\dag}-v^2)^2 \frac{}{}\right )
\end{equation}
where $D_{\mu}\phi=\partial_{\mu}\phi-iA_{\mu}\!\ast\!\phi$. The
gauge field configuration (\ref{potential}) 
with $\phi=0$  still solves the equations of motion but with
 Higgs on the local
maximum of the potential.  One can show that the  noncommutative
vortex solution found in \cite{bak} for spatial noncommutativity
solves the above  1+1 dimensional Abelian-Higgs theory  by
replacing $x^2\rightarrow x^0$ and $A_{2}\rightarrow A_{0}$. Note
that the vortex solution satisfies $D_{1}F_{12}=D_{2}F_{12}=0$,
$D_{\mu}\phi=0$ and $V^{\prime}(\phi)=0$, which makes the equation
of motion  hold. Again these equations are not sensitive to the
signature of the metric.

Solutions localized in spacetime can still  be constructed  if one
adds an adjoint real scalar coupled to the U(1) gauge fields. The
action is similar to (\ref{2d-gauge-fund}) but now with more
general potential
 and the following covariant derivative:
\begin{equation}
  D_{\mu}\phi=\partial_{\mu}\phi-i(A_{\mu}\ast\phi-\phi\ast
  A_{\mu})\,.
\end{equation}
Keeping the gauge field solution as in the previous ones (i.e.
with $\hat F_{10}=\frac{1}{\theta} \hat{P}
$), we wish to solve the equations of motion
\begin{eqnarray}
  &&D_{\mu}\, F^{\mu\nu}\,= -i(\phi\ast D^{\nu}\phi-D^{\nu}\phi\ast
  \phi)\,,\nonumber\\
  && D_{\mu}D^{\mu}\phi =V^{\prime}(\phi)
\ .
\end{eqnarray}
To this end, it suffices to find $\phi$ such that $D_{\mu}\phi=0$
and $V^{\prime}(\phi)=0$. Using the solution (\ref{potential}),
the former equation leads to the following condition,
\begin{equation}
  [\hat S\hat x^\mu \hat S^\dag,\,\,
\hat\phi\ ]=0\,. 
\end{equation}
This is solved by $\hat\phi=\hat\phi_{K
}$
where $\hat\phi_{K}\,\in\, Ker S^\dag$ i.e.
$\hat S^\dag\hat\phi_{K
}=0$. In case of
nonvanishing potential, we have to find the solution of
$V^{\prime}(\phi)=0$ by choosing suitable $\phi_{K
}$,
which is known for a large class of potentials\cite{gms}. If there
is no potential, the above ansatz solves the field equations.
For instance, $\hat F_{10}=\frac{1}{\theta}|0\rangle\langle 0|$
and $\hat\phi=a |0\rangle\langle 0|$ corresponds to such a
solution. Translating into the function  language, it is clear
that the solution again has a Gaussian shape localized in
spacetime.

So far we have discussed various solutions localized in 1+1
dimensional spacetime, but such localized solutions may also exist
in higher dimensional spacetime noncommutative theories.
For example, one may consider 3+1 dimensional
$U(1)$ gauge theory with 
$[x^\mu,x^\nu]=i\theta^{\mu\nu}$. There are substantial
differences depending on the signatures of two Lorentz invariants,
\begin{equation}
  \theta^{\mu\nu}\theta_{\mu\nu}=
2(|\vec\theta_m|^2 -|\vec\theta_e|^2)\ \ ,\
  \
  \epsilon_{\alpha\beta\gamma\delta}
\theta^{\alpha\beta}\theta^{\gamma\delta}=
  8\vec\theta_e\cdot\vec\theta_m \,.
\end{equation}

If $\vec\theta_e\cdot\vec\theta_m\neq 0$, one can make a suitable
Lorentz transformation to set $[\hat x^0, \hat x^1]=i\theta$,
$[\hat x^2,\hat x^3]=-i\theta^\prime$ with all the other pairs
commuting. One can find localized solutions 
 in 3+1 dimensional
spacetime by a slight generalization of the expression
(\ref{potential}):
\begin{equation}\label{3+1-gauge-sol}
  \hat A_{\mu}=-\theta^{-1}_{\mu\nu}
(\hat x^{\nu}-\hat S\hat x^{\nu}\hat S^{\dag}),\ \
  \hat S^\dag\hat S=1,\
  \ \hat S \hat S^\dag=1-\hat P\ .
\end{equation}
Again $\hat P$ is any projection operator, now acting on a Hilbert
space spanned by eigenstates
\begin{equation}
  |m,n\rangle\equiv\frac{(c^\dag)^m}{\sqrt{m!}}\frac{(\tilde c^\dag)^{n}}{\sqrt{n!}}|0,0\rangle
  \ \ ({\rm where}\ c=\frac{x_1-ix_0}{\sqrt{2\theta}},\ \tilde
  c=\frac{x_2-ix_3}{\sqrt{2\theta^\prime}})
\end{equation}
of double harmonic oscillators. 
The above solution solves the equation of motion
(\ref{u(1)-eq-general}) with either choice of the metric;
Lorentzian   or Euclidean. The field strength is now given by
$\hat F_{\mu\nu}=\theta^{-1}_{\mu\nu}\hat P$. With the simple
choice $\hat P = |0,0\rangle\langle 0,0|$, we get the field
strength localized in spacetime as
\begin{equation}
  F_{10}=\frac{4}{\theta}
\exp\left[-\frac{t^2+x^2}{\theta}-
  \frac{y^2+z^2}{\theta^\prime}\right],\
  F_{23}=\frac{4}{\theta^\prime}
\exp\left[-\frac{t^2+x^2}{\theta}-
  \frac{y^2+z^2}{\theta^\prime} \right]\,,
\end{equation}
while other components are zero.

Coupling a real adjoint scalar to the gauge field, again one may
obtain solutions where the matter part is localized in spacetime
too. The procedure is very similar to the 1+1 dimensional case
with adjoint scalar, where the scalar satisfies $D_{\mu}\phi=0$.

For the case $\vec\theta_e\cdot\vec\theta_m=0$, one has to
consider the sign of another invariant.
$\vec\theta_e^2<\vec\theta_m^2$ yields the usual spatial
noncommutativity (i.e. $\vec\theta_e=0$) by suitable Lorentz
transformation, so let us turn to the other two cases.

If $\vec\theta_e^2 > \vec\theta_m^2$, one can make a Lorentz
transformation to the frame where 
$[x_0,x_1]=i\theta$ while other pairs commute. Simply using  the
solution (\ref{e-field}), we get a solution localized in time
direction; a 2-dimensional transient `sheet' occurring at $t=0$.

For the lightlike case $\vec\theta_e^2 = \vec\theta_m^2$
\cite{aha,cai}, one can make a Lorentz transformation to a frame
where only $[x^-,x^2]=i\theta$ is nonzero with $x^{\pm}\equiv x^0
\pm x^1$. We found a large class of solutions to the Maxwell
equation
for this case, but none of them is localized in the lightcone time
$x^+$ direction.

In the solutions constructed above, the fields become
exponentially small as $|t|\rightarrow\infty$. 
These may be
regarded as  spacetime localized degrees arising from the
free vacuum. (See also the related argument in the next
section.) One may also ask if it is possible to find such
a spacetime localized fluctuation that approaches 
asymptotically  a 
scattering
state instead of the vacuum. 
We shall construct such classical solutions, which will verify
explicitly their existence.

First we consider the $1+1$ dimensional $U(1)$ theory. Without
matter, there would be no solution that contains scattering states 
because there is no transverse photon 
in the $1+1$ dimensions. With a real
adjoint scalar, it can be shown that 
there  exist solutions,
\begin{eqnarray}
  &&\hat{A}_{\mu}=-\frac{1}{\theta}\epsilon_{\mu\nu}
  (\hat{x}^{\nu}-\hat{S}\hat{x}^{\nu}\hat{S}^\dag)\ ,
  \nonumber\\
  &&\hat\phi=
\hat{S}\,
{\rm Re}(\phi_0\
  e^{-ip_{\mu}\hat{x}^{\mu}})\,
  \hat{S}^\dag\ ,
\label{1-ptl-2d}
\end{eqnarray}
where $p_{\mu}p^{\mu}=0$ 
and $\phi_0$ is any complex number. The first line of the
solution (\ref{1-ptl-2d}) describes the localized fluctuation
whereas the second line contains a plane wave traveling with
momentum $p^{\mu}$. One may rigorously show that it approaches 
asymptotically a plane wave. For instance, choosing $\hat
S=\sum_{n=0}^\infty|n+1\rangle\langle n|$, we have $\hat
P=|0\rangle\langle 0|$. In functional representation, this is
localized near $x^2=x_0^2+x_1^2<\theta$. Hence, in the asymptotic
region $x\gg\sqrt\theta$, one expects that
$S(x)\,\ast\,S(x)^\ast\,\approx\, 1
$ up to
${\mathcal{O}}(\frac{\theta}{x^2})$ corrections, which implies that
$S(x)$ is unitary  asymptotically. 
 By an explicit 
computation, one finds that
\begin{eqnarray}
  &&S(x)=e^{-i\tan^{\!-1\!}(\frac{x_0}{x_1})}\left(1+
{\mathcal{O}}(\frac{\theta}{x^2})\right)\
  ,\nonumber\\
  &&\phi(x)=S(x)\ast e^{-ip\cdot x}\ast S(x)^\ast=e^{-ip\cdot x}
  \left(1+{\mathcal{O}}(\theta\frac{p\cdot x}{x^2})\right)\ .\nonumber
\end{eqnarray}
It is clear  that (\ref{1-ptl-2d}) is a plane wave for
sufficiently large $x^2 (\, =x_0^2-x_1^2)$. In fact using an appropriate gauge 
transformation, the power-law corrections may be removed 
completely, for the initial surface at a large negative time,
leaving only exponential corrections. 

This kind of solution is not restricted to lower dimensional
theories. For the $3+1$ dimensional $U(1)$ theory, let us
consider the case $\vec\theta_{e}\cdot\vec\theta_m\neq 0$  first. One can
find an exact solution: 
\begin{equation}
  \hat{A}_{\mu}=-\theta^{-1}_{\mu\nu}
  (\hat{x}^{\nu}-\hat{S}\hat{x}^{\nu}\hat{S}^\dag)+
  \hat{S}\,
{\rm Re}(\varepsilon_{\mu} e^{-ip_{\nu}\hat{x}^{\nu}})\,
  \hat{S}^\dag\ ,
\end{equation}
where 
$p_{\mu}p^{\mu}=\varepsilon_{\mu}p^{\mu}=0$. 
This solution describes an asymptotic
  photon with momentum $p_{\mu}$ and
polarization $\varepsilon_{\mu}$. As for 
$\vec\theta_{e}\cdot\vec\theta_m = 0$ with
$\vec\theta_{e}^2>\vec\theta_{m}^2$, we 
consider the case where
only $[x_0,x_1]=i\theta$ is  
noncommutative. It can be shown
that there exists 
a solution
\begin{eqnarray}
  &&\hat{A}_{\mu}=-\frac{1}{\theta}\epsilon_{\mu\nu}
  (\hat{x}^{\nu}-\hat{S}\hat{x}^{\nu}\hat{S}^\dag)+
  \hat{S}\,
{\rm Re}(\varepsilon_{\mu}
  e^{-i(p_{\mu}\hat{x}^{\mu}+p_k x^k)})\,
  \hat{S}^\dag\ ,\nonumber\\
  &&\hat{A}_{j}=\hat{S}\,
{\rm Re}(\varepsilon_j
  e^{-i(p_{\mu}\hat{x}^{\mu}+p_k x^k)})\,
  \hat{S}^\dag\ ,
\label{1-ptl-sheet}
\end{eqnarray}
with $p^2=p_{\mu}p^{\mu}+p_i p_i=0$ and
$p\cdot\varepsilon=p_{\mu}\varepsilon^{\mu}+p_i\varepsilon_i=0$.
(
$\mu,\nu=0,1$ and $i,j=2,3$.) 
This describes a plane wave 
superposed with a spacetime localized sheet centered at $x_0=x_1=0$.

These  examples show that spacetime localized fluctuations 
may arise not just from the free vacuum but also 
from scattering states. 

Finally  let us discuss here about gauge invariant observables for
the 1+1 spacetime noncommutative gauge theories\cite{park}. One
may in fact show that all the gauge invariant variables take the
form of $\int d^2x f(X^\mu)$ where $f$ is an arbitrary
function\cite{park} of the covariant coordinate defined by $X^\mu=
x^\mu +\theta^{\mu\nu}A_\nu$, which transforms covariantly under
the gauge transformation. Then for the configuration of spacetime
localized solutions, any observable quantities that are finite
indicate that the configuration is localized in space and time.
For instance, we define the average position by
$$
q^\mu\equiv {\int d^2x\,\,   X^\mu \,\, F^2_{01} \over \int
d^2x\,\, F^2_{01}}
$$
and the moments\cite{klee}
$$
\Delta^k \equiv {\int d^2x\,\, (X^\mu-q^\mu)^k \,\, F^2_{01} \over
\int d^2x\,\, F^2_{01}}
$$
where we average over the action density. On the spacetime
localized solutions, one may show that all the higher moments ($k
\ge 2$) vanish. Thus we conclude the configurations are centered
at $q^\mu$ and  localized indeed.


\section{Discussions}

In this note, we construct Lorentzian solutions of spacetime
noncommutative gauge theories, which are exponentially localized
in time as well as space. Together with the time
translational invariance, we argue that 
the perturbative  S matrix based on the free field degrees is
problematic. In the 3+1 dimensional spacetime noncommutative gauge
theories, the problem disappears if $\vec{\theta}_e\cdot
\vec{\theta}_m=0$ and $ \vec{\theta}_e^2\le
\vec{\theta}_m^2$. 
Otherwise there exist  spacetime localized Lorentzian solutions in
3+1 dimensional gauge theories.

The situation here is similar to the case of tachyonic
instabilities in quantum mechanical system or field theories. To
illustrate the point,  consider a dynamical system described by
$L={1\over 2} {\dot{x}^2}-V(x)$ with a potential
\begin{eqnarray}
V(x)={1\over 2}(-{x^2
} + {x^4
})
\end{eqnarray}
There is a solution $x(t)= \pm\,\, {\rm sech} (t- t_0)$ with the
velocity $v(t)= \mp\,\, {\rm sech} (t-t_0) \tanh(t-t_0)$
 localized exponentially
in time. Existence of such solution does not imply that the
dynamical system is problematic. Rather it represents a tachyonic
instability at $x=0$. Field theories expanded around tachyonic
vacuum have a similar problem. For instance, in  free field
theories with negative mass squared, the norm of certain states is
not preserved and leads to  unitarity problems. Some states have a
time evolution $e^{-iEt}|E\rangle$ with a purely imaginary energy
$E=\pm i\sqrt{|m^2|-\vec{k}^2} $ for $\vec{k}^2 < |m^2|$ and norms
cannot be preserved.

Likewise, the existence of spacetime localized solutions for the
spacetime noncommutative field theories seems
 related to
similar phenomena.
Potential instabilities involved here
 may be argued   in the following way. Using the Seiberg-Witten
map\cite{seiberg}, the spacetime noncommutative (gauge) field
theories may be mapped  to the equivalent system but without
noncommutativity. There the system has a corresponding  background
electric
field. 
Instabilities may arise due to this background electric field.
In this sense, we suspect that our spacetime localized degrees may
be closely  connected to the phenomena of tachyon condensation\footnote{See 
Ref. \cite{al} for the related discussions.} or
S branes\cite{gutperle}. If this is the case, one may further ask
about the true vacuum after tachyon condensation.

Despite attractiveness of such interpretation, however, it should
be remembered that we do not know how to make sense of the
spacetime noncommutative field theories as  quantum mechanics.
In summary, the perturbative  S-matrix based on
 free in and out states
misses  degrees of the spacetime localized solutions.
To settle the further issues, we need a nonperturbative
quantization of the spacetime noncommutative field theories.
Further investigation is necessary in this direction.

In this note, we focus our investigation on the gauge theories,
which are normally the ones that can be obtained from string
theories in the decoupling limit. But as a separate issue, one
could ask the existence of spacetime localized solution for the
scalar theories with potential. Consider for example 1+1
dimensional scalar theory with potential $V(\phi)=- {1\over
2}\phi^2 + {1\over 4}\phi^4$. This is question about whether there
exists GMS (Gopakumar-Minwalla-Strominger) type solution
\cite{gms} but one of spatial coordinate is replaced by time. In
the large $\theta$ limit, one could ignore the kinetic term as in
the GMS case and, for example,
 $\phi= 2 \exp\left[{-{t^2+ x^2\over
 \theta}}\right]$
will be solution in the limit. However unlike the case of the GMS
soliton, if one tries to perturb the above solution for large but
finite $\theta$, one may show that the first order perturbative
solution cannot be solved. This happens quite generically for
other forms of large $\theta$ solution and other forms of
potential. We do not know in this pure scalar case whether there
exist spacetime localized solutions or not. Further investigation
is required in this direction.


%
%

\section*{Acknowledgments}
 We are grateful to   Matthew Kleban, Choonkyu Lee,
Lenny Susskind, Takao Suyama, Tsukasa Tada, Hyun-Seok Yang and
Ho-Ung Yee for useful discussions and conversations. D.B. is
supported in part by KOSEF ABRL R14-2003-012-01002-0 and KOSEF
R01-2003-000-10319-0. S.K. is supported in part by BK21 project of
the Ministry of Education, Korea, and the Korea Research
Foundation Grant 2001-015-DP0085.


\begin{thebibliography}{99}


\bibitem{gomismehen}
J.~Gomis and T.~Mehen, ``Space-time noncommutative field theories
and unitarity,'' Nucl.\ Phys.\ B {\bf 591}, 265 (2000)
[arXiv:hep-th/0005129].

%



\bibitem{aha}
O.~Aharony, J.~Gomis and T.~Mehen, ``On theories with light-like
noncommutativity,'' JHEP {\bf 0009}, 023 (2000)
[arXiv:hep-th/0006236].



\bibitem{al}
L.~Alvarez-Gaume, J.~L.~F.~Barbon and R.~Zwicky,
``Remarks on time-space noncommutative field theories,''
JHEP {\bf 0105}, 057 (2001)
[arXiv:hep-th/0103069].


\bibitem{bahns}
D.~Bahns, S.~Doplicher, K.~Fredenhagen and G.~Piacitelli, ``On the
unitarity problem in space/time noncommutative theories,'' Phys.\
Lett.\ B {\bf 533}, 178 (2002) [arXiv:hep-th/0201222].


\bibitem{rim}
C.~H.~Rim and J.~H.~Yee, ``Unitarity in space-time noncommutative
field theories,'' arXiv:hep-th/0205193.







\bibitem{torr}
A.~Torrielli, ``Cutting rules and perturbative unitarity of
noncommutative electric-type field theories from string theory,''
Phys.\ Rev.\ D {\bf 67}, 086010 (2003) [arXiv:hep-th/0207148].











\bibitem{ohl}
T.~Ohl, R.~Ruckl and J.~Zeiner, ``Unitarity of time-like
noncommutative gauge theories:
 The violation of Ward identities in time-ordered perturbation theory,''
arXiv:hep-th/0309021.


\bibitem{GG}
J.~Gomis, K.~Kamimura and J.~Llosa,
``Hamiltonian formalism for space-time non-commutative theories,''
Phys.\ Rev.\ D {\bf 63}, 045003 (2001)
[arXiv:hep-th/0006235].

\bibitem{CH}
T.~C.~Cheng, P.~M.~Ho and M.~C.~Yeh,
``Perturbative approach to higher derivative and nonlocal theories,''
Nucl.\ Phys.\ B {\bf 625}, 151 (2002)
[arXiv:hep-th/0111160].



\bibitem{polychronakos}
A.~P.~Polychronakos, ``Flux tube solutions in noncommutative gauge
theories,'' Phys.\ Lett.\ B {\bf 495}, 407 (2000)
[arXiv:hep-th/0007043].



\bibitem{bak}
D.~Bak, ``Exact multi-vortex solutions in noncommutative
Abelian-Higgs theory,'' Phys.\ Lett.\ B {\bf 495}, 251 (2000)
[arXiv:hep-th/0008204].





\bibitem{aganagic}
M.~Aganagic, R.~Gopakumar, S.~Minwalla and A.~Strominger,
``Unstable solitons in noncommutative gauge theory,'' JHEP {\bf
0104}, 001 (2001) [arXiv:hep-th/0009142].


\bibitem{harvey}
J.~A.~Harvey, P.~Kraus and F.~Larsen, ``Exact noncommutative
solitons,'' JHEP {\bf 0012}, 024 (2000) [arXiv:hep-th/0010060].


\bibitem{gross}
D.~J.~Gross and N.~A.~Nekrasov, ``Solitons in noncommutative gauge
theory,'' JHEP {\bf 0103}, 044 (2001) [arXiv:hep-th/0010090].


\bibitem{klee}
D.~Bak, K.~M.~Lee and J.~H.~Park, ``Noncommutative vortex
solitons,'' Phys.\ Rev.\ D {\bf 63}, 125010 (2001)
[arXiv:hep-th/0011099].



\bibitem{yee}
D.~Bak, S.~K.~Kim, K.~S.~Soh and J.~H.~Yee, ``Noncommutative
Chern-Simons solitons,'' Phys.\ Rev.\ D {\bf 64}, 025018 (2001)
[arXiv:hep-th/0102137];
D.~Bak and K.~M.~Lee, ``Noncommutative supersymmetric tubes,''
Phys.\ Lett.\ B {\bf 509}, 168 (2001) [arXiv:hep-th/0103148].


\bibitem{park}
D.~Bak, K.~M.~Lee and J.~H.~Park, ``Comments on noncommutative
gauge theories,'' Phys.\ Lett.\ B {\bf 501}, 305 (2001)
[arXiv:hep-th/0011244].



\bibitem{gms}
R.~Gopakumar, S.~Minwalla and A.~Strominger, ``Noncommutative
solitons,'' JHEP {\bf 0005}, 020 (2000) [arXiv:hep-th/0003160].


\bibitem{cai}
R.~G.~Cai and N.~Ohta, ``Lorentz transformation and light-like
noncommutative SYM,'' JHEP {\bf 0010}, 036 (2000)
[arXiv:hep-th/0008119].




\bibitem{seiberg}
N.~Seiberg and E.~Witten, ``String theory and noncommutative
geometry,'' JHEP {\bf 9909}, 032 (1999) [arXiv:hep-th/9908142].


\bibitem{gutperle}
M.~Gutperle and A.~Strominger, ``Spacelike branes,'' JHEP {\bf
0204}, 018 (2002) [arXiv:hep-th/0202210].


\end{thebibliography}
\end{document}